\documentclass{article}
\usepackage{graphicx}
\usepackage{amssymb}
\usepackage{times}
\usepackage{emulateapj}
\usepackage{psfig}

\def\spose#1{\hbox to 0pt{#1\hss}}

\def\multleft#1{\hbox to size{\vbox {\halign {\lft{##}\cr #1}}\hfill}\par}
\def\multright#1{\hbox to size{\vbox {\halign {\rt{##}\cr #1}}\hfill}\par}

\def\boxit#1{\vbox{\hrule\hbox{\vrule\kern3pt\vbox{\kern3pt
          #1 \kern3pt}\kern3pt\vrule}\hrule}}

\begin{document}

\title{A variable efficiency for thin disk black hole accretion}

\author{Christopher~S.~Reynolds}

\affil{JILA, Campus Box 440, University of Colorado,
Boulder CO~80309\altaffilmark{1}}

\and
 
\author{Philip~J.~Armitage}

\affil{School of Physics and Astronomy, University of St.Andrews,
Fife, KY16~9SS, UK.}

\altaffiltext{1}{Present address: Dept. of Astronomy, University of Maryland, College Park, MD~20742}

\begin{abstract}
We explore the presence of torques at the inner edges of geometrically-thin
black hole accretion disks using 3-dimensional
magnetohydrodynamic (MHD) simulations in a pseudo-Newtonian potential.
By varying the saturation level of the magnetorotational instability 
that leads to angular momentum transport, we show that the dynamics 
of gas inside the radius of marginal stability varies depending upon 
the magnetic field strength just {\em outside} that radius. Weak 
fields are unable to causally connect material within the plunging region 
to the rest of the disk, and zero torque is an approximately correct 
boundary condition at the radius of marginal stability. Stronger fields, 
which we obtain artificially but which may occur physically within more 
complete disk models, are able to couple at least some parts of the 
plunging region to the rest of the disk. In this case, angular momentum 
(and implicitly energy) is extracted from the material in the plunging 
region. Furthermore, the magnetic coupling to the plunging region can 
be highly time dependent with large fluctuations in the torque at the 
radius of marginal stability.  This implies 
varying accretion efficiencies, both across systems and within a given system 
at different times. The results suggest a possible link between changes in 
X-ray and outflow activity, with both being driven by transitions between 
weak and strong field states.
\end{abstract}

\keywords{accretion, accretion disks -- black hole physics -- MHD --
 hydrodynamics -- instabilities}

\section{Introduction}

In a wide class of black hole systems, including QSOs, Seyfert galaxies,
and high-state Galactic Black Hole Candidates, the accretion disk is almost
certainly geometrically thin and radiatively efficient.  The radiation
emitted from an annulus of such a disk does not simply equal the change in
gravitational binding energy as gas flows across the annulus, but rather
includes a contribution from energy transported into the annulus from
elsewhere in the disk. As a consequence, the disk structure, along with
related quantities such as the radiative efficiency (the fraction of rest
mass energy which is radiated during the accretion process), depends upon
the location and nature of the boundary conditions at the inner disk edge.

For a black hole accretion disk, a natural location to place the inner
boundary condition is at the radius of marginal stability, $r=r_{\rm
ms}$. This is the radius inside of which circular orbits are no longer
stable, and is at $6GM/c^2$ for a non-rotating (Schwarzschild) black
hole. In the standard model for black hole accretion (Novikov \& Thorne
1973; Paczynski \& Bisnovatyi-Kogan 1981; Abramowicz \& Kato 1989;
Paczynski 2000), one assumes that there is no angular momentum transport
across $r=r_{\rm ms}$. The motivation for this zero torque boundary
condition (ZTBC) is that the radial flow rapidly becomes supersonic once
inside $r=r_{\rm ms}$, and so the material loses causal contact with the
rest of the disk. Thereafter, material simply spirals ballistically into
the black hole. For this reason, we will refer to the region $r<r_{\rm ms}$
as the plunging region.

The ZTBC limits the accretion efficiency to around 6\% (for a non-rotating
black hole) -- already a substantial figure. However, it is possible that
even higher efficiencies may arise. The suggestion is that magnetic fields,
which are generated in the disk by the magnetorotational instability
(Balbus \& Hawley 1991) may become dynamically significant within the
plunging region. These fields could then causally connect the plunging
region to the rest of the accretion disk, allowing energy and angular
momentum to be extracted from gas as it executes its final spiral into the
black hole (Krolik 1999; Gammie 1999; Agol \& Krolik 2000). Subsequent
numerical simulations (none of which are General Relativistic) have
confirmed the presence of magnetic torques at the marginally stable orbit,
but have reached different conclusions as to their impact on the dynamics
of the flow.  Global MHD simulations of adiabatic accretion tori (Hawley
2000; Hawley \& Krolik 2001) showed outward angular momentum transport
continuing within $r_{\rm ms}$, while unstratified simulations suggested
more modest effects (Armitage, Reynolds \& Chiang 2001; Hawley 2001). The
important parameters or numerical differences that cause this different
behavior remain to be determined.

In this {\it letter} we explore the suggestion (Charles Gammie, private
communication) that the dynamics of the flow within the
plunging region may depend upon the strength of magnetic fields 
in the disk at $r > r_{\rm ms}$. We extend our earlier work by 
presenting `boosted' simulations in which the strength of the magnetic 
field in the saturated turbulence is enhanced by starting with an initially 
large seed field. This is a numerical trick, though 
it may have a physical counterpart if the inner disk is 
threaded by a large-scale magnetic field. For the purposes of 
this letter, however, the aim is solely to allow us to 
study how the dynamics of the plunging region vary with field 
strength in a controlled fashion. We show that 
the dynamics of gas in the plunging region, and the validity 
of the ZTBC, do indeed depend upon the saturation field strength, 
and discuss the implications for observations of 
accreting black holes.

\section{The simulations}

\begin{table*}
\begin{center}
\begin{tabular}{ccccccc}
Run & $\phi_{\rm max}$ & $z_{\rm max} / r_{\rm Sch}$ & $n_r$ & $n_\phi$ & $n_z$ &
$\beta_{\rm i}$ \\\hline
1  &  $\pi/4$ & $0.5$ & $200(40)$ & $60$ & $40$ & $5000(z)$ \\
2  &  $\pi/4$ & $0.5$ & $200(40)$ & $60$ & $40$ & $500(z)$ \\
3  &  $\pi/4$ & $0.5$ & $200(40)$ & $60$ & $40$ & $100(\phi)$ \\
4  &  $\pi/2$ & $0.5$ & $200(40)$ & $120$ & $40$ & $5000(z)$ \\\hline
\end{tabular}
\end{center}
\caption{The set of simulations.  Columns show (1) Run number, (2)
$\phi$-domain, (3) $z$-domain, (4) number of radial cells (and number
inside $r=r_{\rm ms}$), (5) number of $\phi$ cells, (6) number of vertical
cells, and (7) initial $\beta$ of plasma (and direction).}
\end{table*}

We use the ZEUS code to solve the equations of ideal MHD (Stone \&
Norman 1992a, 1992b; Clarke, Norman \& Fielder 1994; Norman 2000) in
cylindrical co-ordinates, using a setup very similar to that described
previously (Armitage, Reynolds \& Chiang 2001). Stated briefly, the
computational domain is the wedge given by $r\in(r_{\rm in},r_{\rm
out})$, $\phi\in(0,\phi_{\rm max})$ and $z\in(-z_{\rm max},z_{\rm
max})$. The effects of General Relativity (in particular the existence
of an innermost stable orbit) are modelled in this inherently
non-relativistic code by using a pseudo-Newtonian potential (Paczynski
\& Wiita 1980), $\Phi=-\frac{GM}{r-r_{\rm Sch}}$, where $r_{\rm
Sch}=2GM/c^2$. The vertical component of gravity is neglected, so the
disk has no vertical structure. There is therefore no time-averaged
variation of density or magnetic field strength with $z$.  The
equation of state is isothermal, with a sound speed $c_{\rm s}$ which
is assumed uniform and constant. At $r_{\rm ms}$, the ratio of the
sound speed to the Keplerian velocity is $c_s / v_\phi = 0.065$,
slightly cooler than our previous simulations.

The initial conditions for density and velocity were obtained by 
relaxing a Gaussian radial density profile, using ZEUS in its 1D
hydrodynamic mode, to produce a density profile which is in
accurate numerical equilibrium.  The initial configurations of the 
3D MHD simulations were then produced by adding a seed magnetic field.
Table~1 shows the computational domain, resolution, initial value of 
$\beta$, and seed field configuration for the simulations. As
usual, $\beta$ is defined as the ratio of thermal energy density to
magnetic energy density. In all cases, we set $r_{\rm in}=4GM/c^2$ and
$r_{\rm out}=20GM/c^2$. The radial boundary conditions were such as to
permit outflow, and all other boundaries were made periodic. These
simulations were then evolved until the magnetorotational instability (MRI)
became non-linear and produced fully developed MHD turbulence 
(in the sense that the MRI has saturated and there is no further 
long-term growth of the magnetic energy).  This turbulence drives 
accretion of material through the inner radial boundary, and the run is 
continued until an appreciable amount of material has left
the computational domain..

\section{The dynamics of the flow within the plunging region}

\begin{figure*}[t]
\hbox{
\psfig{figure=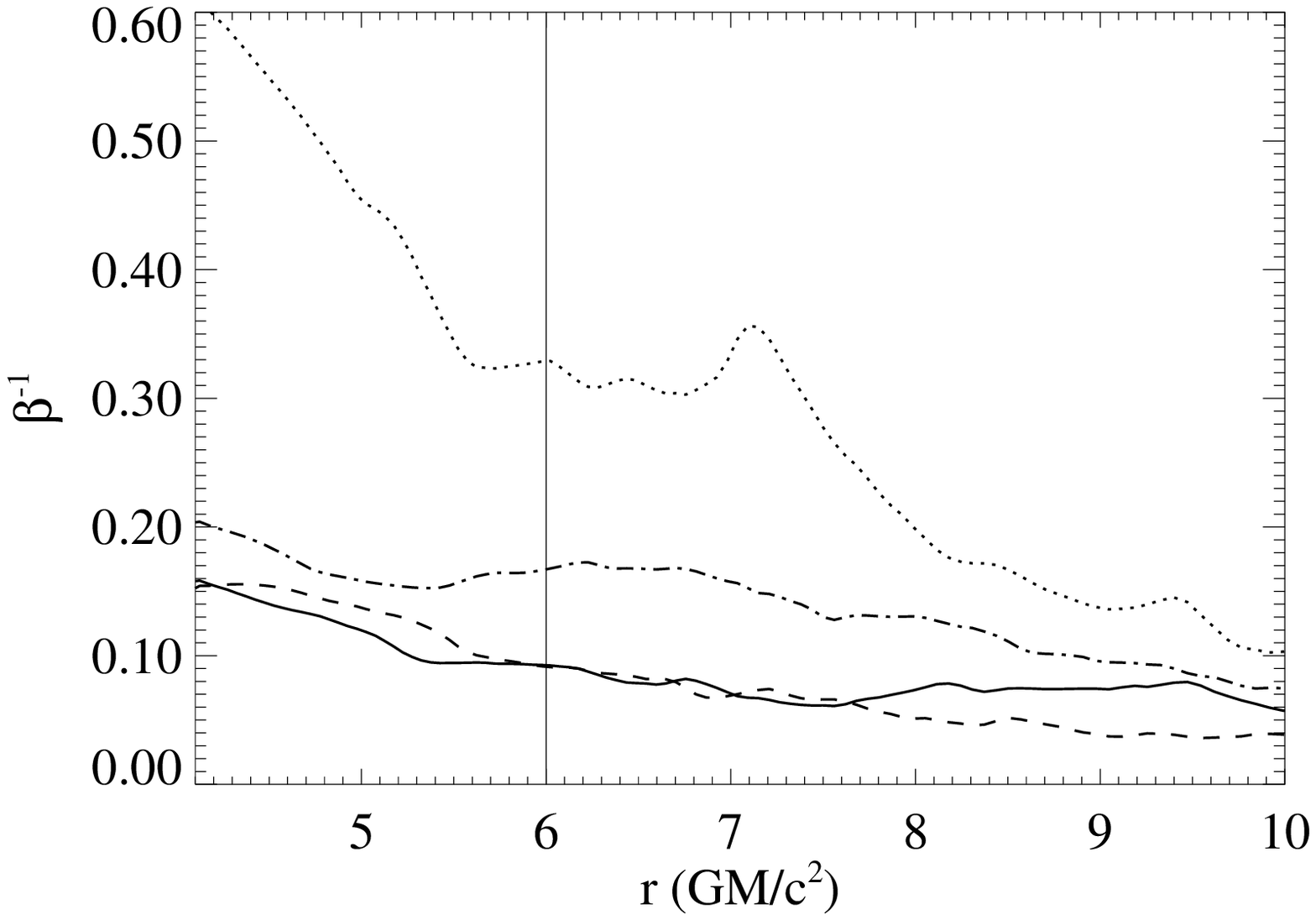,width=0.48\textwidth}
\psfig{figure=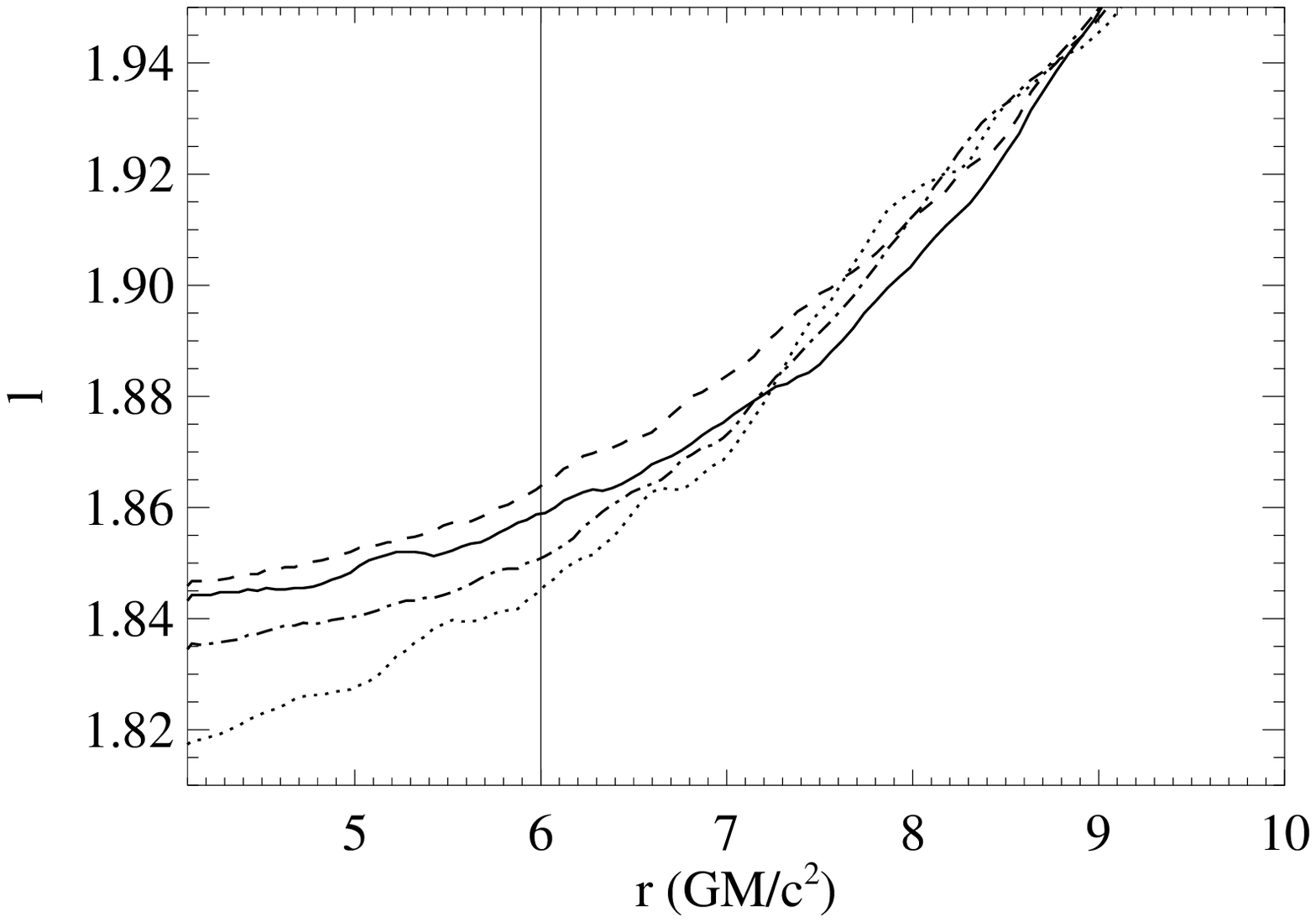,width=0.48\textwidth}
}
\caption{Time-averaged radial profiles of $\beta^{-1}$ (the ratio of
magnetic energy density to thermal energy density) and $l$ (specific
angular momentum) for our high resolution simulations.  The runs are
denoted by: Run~1 (solid line), Run~2 (dotted line), Run~3 (dashed line),
Run~4 (dot-dashed line).}
\end{figure*}

\begin{figure*}[t]
\hbox{
\psfig{figure=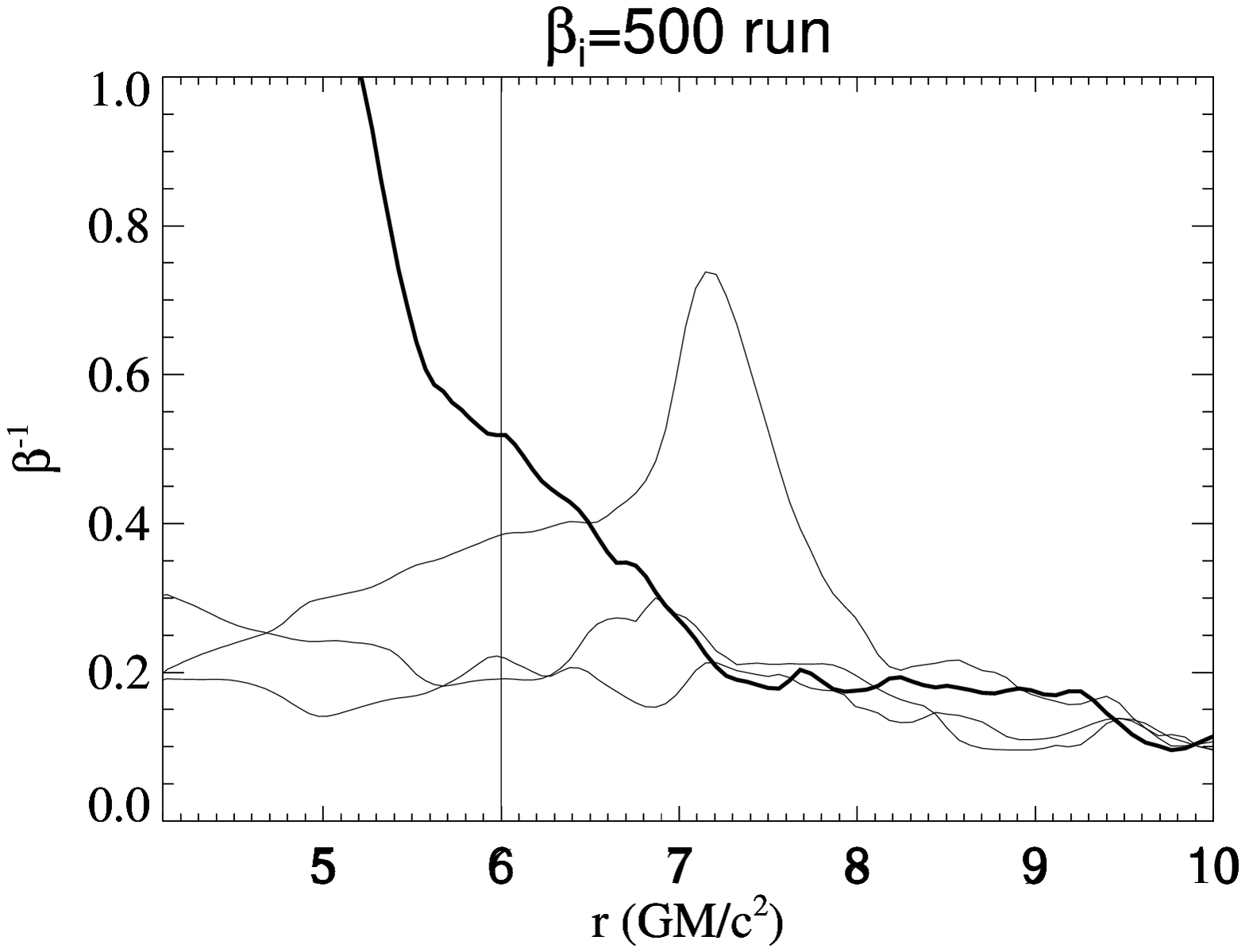,width=0.48\textwidth}
\psfig{figure=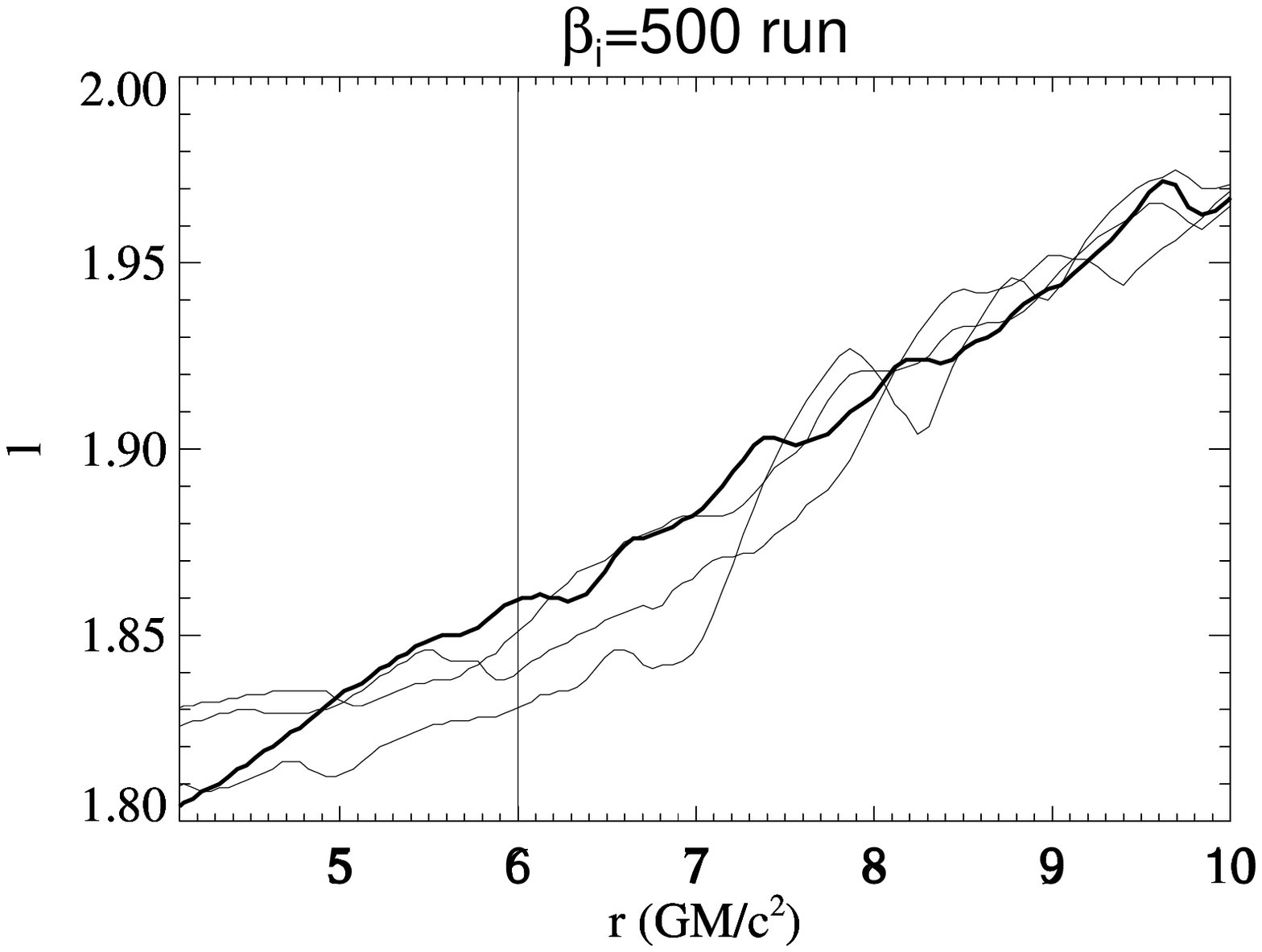,width=0.48\textwidth}
}
\caption{Radial profiles of $\beta^{-1}$ and $l$ for individual timeslices
from the boosted simulation, Run~2.  In particular, note the anomalous
event (shown with a thick line) during which the plunging region becomes
magnetically dominated and there is efficient angular momentum transport
out of this region.}
\end{figure*}

\subsection{Time averaged properties}

We consider first the time-averaged properties of our runs.
Figure~1 shows time-averaged radial profiles for $\beta^{-1}$ (i.e.,
the ratio of magnetic field energy density to thermal energy
density) and the specific angular momentum $l$. As intended, the 
magnetic field in Run~2, which starts with a relatively strong vertical 
seed field ($\beta_z = 500$), saturates with significantly higher 
relative magnetic field energy than those runs which start with either 
a weaker initial field or a comparable toroidal field.  This is due to 
the continued boosting of the MRI from the conserved vertical flux 
that accompanies the initial vertical field (Hawley, Gammie \& 
Balbus 1995). Increasing the size of the computational domain also 
increases the saturation value of the magnetic field, but by a 
smaller factor. 

The dynamics of the flow -- diagnosed using the specific angular
momentum -- correlates with the saturation field strength.  Firstly, 
consider the unboosted runs (Runs~1,3 and 4; shown by the solid, dashed 
and dot-dashed lines respectively).  As one approaches and enters the 
plunging region, the radial profile of $l$ flattens. 
If the ZTBC is strictly valid, $l$ should be constant within the
plunging region. For these runs, we obtain the same result as 
before (Armitage, Reynolds \& Chiang 2001), namely that the ZTBC yields a 
good but not perfect description of the inner disk dynamics. Indeed, 
examination of the $r\phi$ component of the magnetic stress tensor 
shows that the Maxwell stresses at $r=r_{\rm ms}$ do not vanish but 
are less than the peak stress, which occurs at $r\sim 10-12\,GM/c^2$, 
by a factor of 5--8.

On the other hand, angular momentum transport within the plunging region is
much more significant in the boosted run (Run~2; dotted line in Fig.~1).
The time averaged stress at $r=r_{\rm ms}$ is still less than the peak value
(which occurs at $r\sim 8\,GM/c^2$), but only by a factor of 2. 
Significant stresses continue to operate within the region $r<r_{\rm ms}$, 
and there is evidently a clear violation of the ZTBC.

\subsection{Variability within individual simulations}

Turbulent accretion disks are inherently variable systems and so it is
also interesting to examine variability during our simulations.  For
this purpose, we will compare and contrast Run~1 and Run~2.  Both runs
display order unity temporal and spatial fluctuations in $\beta^{-1}$.
For the unboosted run these fluctuations have only a small effect on
$l$. For the boosted run shown in Fig.~2, however, there are much
larger fluctuations in $l$. In particularly, during one timeslice from
Run~2, a large part of the plunging region becomes magnetically
dominated ($\beta<1$).  During this period (shown with a thick line in
the bottom panels of Fig.~2) $l$ declines within the plunging region
with the same slope as in the rest of the disk, with the Maxwell
stresses staying fairly constant well into the plunging region.
Events as dramatic as this are fairly rare (caught in 1 out of the 12
independent timeslices that we analyzed), but it is clear that the
stress at $r=r_{\rm ms}$ is a rapidly varying function of time, with
much more frequent occurrences of moderate stress.

Extraction of energy and angular momentum from within the plunging
region can only occur if the material can remain causally connected to
the rest of the disk. Analysis of the simulations shows that large
stresses occur within the plunging region when the magnetic fields are
strong enough for the {\it peak} radial MHD wave speed to exceed the
radial inflow velocity throughout a large part of the plunging region.
It is worth noting that even during the magnetically dominated event
seen in Run~2, the {\it azimuthally averaged} MHD wave speeds are
always slower that the inflow speeds.  The angular momentum transport
within this highly inhomogeneous plasma therefore appears to be
mediated by low-$\beta$ (i.e., higher relative magnetic field)
filaments.

\section{Discussion}

\subsection{Radiative efficiency}

Calculating the implied change to the radiative efficiency as 
a consequence of the magnetic torque at $r_{\rm ms}$ is not 
straightforward (see the discussion in Hawley \& Krolik 2001). 
A simple approach is to note that the dissipation rate is
\begin{equation} 
 Q(r) = \vert { {{\rm d} \Omega} \over {{\rm d} \ln r } } \vert 
 \int T^{r \phi} dz,
\end{equation} 
where $T^{r \phi}$ is the local stress. We can then estimate the 
change in the radiative efficiency by comparing $Q(r)$, computed 
using the stress measured in the simulations, with the form 
expected in a steady disk if there were no stress at $r_{\rm ms}$. 
 
In practice, the non-steady nature of the flow, and the 
limited radial extent of the simulations, creates 
difficulties. An indicative estimate, however, follows from noting 
that in run 2 the mean magnetic stress is roughly constant 
between $r_{\rm ms}$ and $2 r_{\rm ms}$. Compared to a 
standard disc model, this implies a 50\% increase in 
the dissipation in the region $r_{\rm ms} < r < 2r_{\rm ms}$, 
and a 10\% increase over $r_{\rm ms} < r < 10r_{\rm ms}$. 
Significantly larger effects are seen in some individual 
timeslices.

\subsection{Comparison with previous work}

The simulations presented here suggest that the different 
dynamics within the plunging region, obtained by Hawley \& 
Krolik (2001) and Armitage, Reynolds \& Chiang (2001), are 
plausibly due to different values in the saturation levels 
of magnetic fields in the disk. The relatively strong 
magnetic fields in the former simulations, which obtained $\beta\sim 5-10$, 
are more comparable to the boosted simulations in the current 
work, and lead to substantial angular momentum transport within the 
plunging region. Weaker disk fields, with $\beta\sim 20$ and 
a Shakura-Sunyaev (1973) $\alpha$ parameter of $\alpha\sim
0.02-0.04$, produce much less striking effects. Although there 
are undoubtably other significant differences between the 
simulations, all work to date is broadly consistent with 
the hypothesis that significant magnetic coupling to 
the plunging region requires disk fields with 
$\alpha\sim 0.1$ (and $\beta\sim 10$). This applies 
for disks that are moderately thin, with relative 
thickness $h/r\sim 0.1$, which are the only systems that 
have been simulated so far.

\subsection{Scaling to thinner disks}

Very thin disks are difficult to simulate, and it is not obvious how
the threshold $\beta$ seen in the current work scales with $h/r$. We
note, however, that the Alfven speed in the disk scales as, $v_A
\propto \left( {h \over r} \right) \beta^{-1/2}$.  Unless $\beta$ in
the disk decreases rapidly with decreasing $h/r$, the Alfven speed in
the gas immediately outside the plunging region will decrease with
decreasing disk thickness. Conversely, the inflow speed $v_r$ within
the plunging region is a fixed function of $r$, and is independent of
$h/r$. Hence, a larger increase in the Alfven speed is needed if
thinner disks are to maintain causal contact into the plunging
region. It is therefore probable that it becomes increasingly more
difficult to extract energy from the plunging region as the disk
becomes thinner, although further simulation work is required before
drawing robust conclusions.

\section{Conclusions}

In this letter, we have used MHD simulations to investigate the effect
of the magnetic field strength in the body of the accretion disk on
the dynamics of the material within the plunging region. For a disk
with a sound speed corresponding to an $h/r \approx 0.1$, we find
that plausible saturation values of the magnetic field are within a
factor of a few of a threshold between two regimes. In the runs which
saturated with a fairly low magnetic field ($\beta\sim 20$), there was
only modest extraction of angular momentum from material within the
plunging region, and the ZTBC was a reasonable approximation.  In our
`boosted' runs, where the field saturated at a higher level
($\beta\sim 5-10$), we found significantly stronger angular momentum
transport within the plunging region.  Furthermore, our boosted run
showed sporadic intervals of very efficient angular momentum transport
(with stresses that remained fairly constant well within the plunging
region). We note that stratified simulations by other authors (Hawley
2000; Miller \& Stone 2000; Hawley \& Krolik 2001) suggest that
magnetic fields could be stronger (i.e. lower $\beta$) above the disk
midplane. Such fields, which are not included in our simulations,
could contribute to more significant transport within the plunging
region.

The extraction of energy and angular momentum from material in the
plunging region can significantly increase the efficiency of accretion
above the standard value, especially in the innermost regions. This
raises the possibility of varying efficiencies, both across systems,
and within a given system at different times.

In this paper, we have used a large scale flux threading the accretion
disk as a numerical device for altering the saturation level of the
MRI. Large scale fields could also, of course, be present in real
systems, and are often invoked in models for the formation of jets and
outflows (e.g. Ouyed, Pudritz \& Stone 1997, and references therein).
This raises the interesting possibility that periods of increased
luminosity (due to a high radiative efficiency) may be related to
periods during which strong outflows occur.  Such associations have
been reported for Galactic Black Hole Candidates such as GRS~1915+105
(Mirabel et al. 1998).

\section*{Acknowledgments}

We extend a special thanks to Charles Gammie for suggesting this
problem.  We are grateful to the developers of ZEUS and ZEUS-MP for
making these codes available as community resources.  CSR acknowledges
support from Hubble Fellowship grant HF-01113.01-98A.  PJA thanks JILA
for hospitality.

\end{document}